\documentclass[pre,superscriptaddress,onecolumn,a4paper,showpacs,notitlepage]{revtex4-1}
\usepackage{graphics}
\usepackage{amssymb}
\usepackage{wasysym}
\usepackage{latexsym}
\usepackage{graphicx}
\usepackage{epsfig}
\usepackage{subfigure}

\begin{document}

\title{The synergy between two threats: disinformation and Covid-19}

\author{Henrique A. T\'ortura}
\affiliation{Instituto de F\'{\i}sica de S\~ao Carlos,
  Universidade de S\~ao Paulo,
  Caixa Postal 369 \\ 
  S\~ao Carlos, S\~ao Paulo   13560-970, Brazil}

\author{Jos\'e F. Fontanari}

\affiliation{Instituto de F\'{\i}sica de S\~ao Carlos,
Universidade de S\~ao Paulo,
  Caixa Postal 369 \\ 
  S\~ao Carlos, S\~ao Paulo   13560-970, Brazil }

\begin{abstract}
The breakdown of trusted sources of information is probably  one of the most serious problems today, since in the absence of a common ground, it will be impossible to address   the problems that trouble our contemporary world. The Covid-19 pandemic is  just  a recent   situation  where the lack of agreed stances has led  to failure and hopelessness. In fact, disinformation surrounding the Covid-19  has  been a distinctive feature of this pandemic since its very beginning and  has  hampered what is perhaps  the most important initiative to prevent the spread of the coronavirus, viz., an effective communication between scientifically-minded health authorities and the general public.  To investigate how disinformation threatens  epistemic security,  here we propose and solve analytically an evolutionary game-theoretic model where   the individuals must accurately estimate  some property of their  hazardous environment.  They can either explore the  environment or copy the estimate from another individual, who may   display a distorted version of its estimate.   We find  that the 
exploration-only strategy is optimal  when the  environment is relatively safe  and the individuals are not reliable. In this doomsday scenario, disinformation  erodes trust and  suppresses the  ability  of the individuals to  share information with one another.
\end{abstract}

\maketitle

\section{Introduction}	

The Covid-19 pandemic has impacted our society in many different forms, some of which will probably only  be known many years from now. In view of these broad repercussions, it is desirable that  the mathematical and computational  studies of the Covid-19 pandemic  go beyond  the analysis of the  transmission of the physical infection through standard epidemiological models.  The understanding of the more nuanced effects of the disease in the distinct age and socioeconomic  segments of the population requires a multiscale  and multifaceted approach.\cite{Aguiar_2021,Bellomo_2020,Bellomo_2022,Bertozzi_2020}  This approach should consider also  the impact   of the   measures to curb the spread of coronavirus (e.g., social distancing and lockdowns)  on the  mental health of the population,\cite{Liang_2020,Miller_2020,Fontanari_2021}  and on  work productivity.\cite{Kitagawa_2021,Peter_2021,Chapter_2021} This impact  can be thought of as a second-order effect of the virus infection. Interestingly, the surge of loneliness associated to those prevention measures  may be related to the widespread rising of authoritarianism seen today. In fact,  as pointed out long ago by  Hannah Arendt, ``the chief characteristic of the mass man is not brutality and backwardness, but his isolation and lack of normal social relationships''.\cite{Arendt_1973} Thus, the present  retrogress to authoritarianism could well be seen as a third-order effect of the coronavirus.

Perhaps one of the most important initiatives to prevent the spread of the coronavirus is an effective communication between the  health authorities and the public, so as to   guarantee universal access to scientific-based content about the disease. However, misinformation surrounding the Covid-19  has  been a distinctive feature of this pandemic since its very beginning in Wuhan in late 2019.\cite{Maxmen_2022} In a sample of statements about the coronavirus rated false or misleading by fact-checkers between January and March 2020, it was found that 59\% of misinformation was created by reconfiguring and recontextualizing information and 38\% was entirely fabricated.\cite{Reuters_2020} Here the   term  misinformation stands for inaccurate information that can can mislead and harm people, regardless of whether it results from an honest mistake, negligence,  or  intentional deception.  Only in the latter  case we use the term  disinformation, which is then misinformation with the explicit intention to mislead.\cite{Fallis_2014,Fallis_2015}
In addition,  disinformation can  erode trust and  inhibit our ability to  share information with one another, i.e., disinformation  is a threat to epistemic security. In fact, the breakdown of trusted sources of information is probably one of the most pressing problems today.\cite{Seger_2020}

In this contribution we address quantitatively the threat of  disinformation to  epistemic security using an individual-based model where the individuals  must accurately estimate  a property of their surroundings in order to survive the environmental challenges. Only the survivors  have a chance to contribute offspring to future generations.  The individuals  can either explore their  surroundings or copy the estimate made by another individual.  The  hazardousness of the environment is determined by the parameter $\sigma^2$.
The  copying probability $w$, which we assume to be  the same for all individuals,  is a proxy for their credulity  or, equivalently,  for the degree of trust of the community.  This is the leading parameter of  the model and our aim is to determine the value  $\tilde{w}$ that maximizes the probability of survival of the individuals.  In fact, the broad view of culture as something that we learn from each other\cite{Kroeber_1952,Blackmore_2000} builds on  copying (or sharing) of  information, which suggests use of $\tilde{w}$ as a measure of the degree of epistemic security of a community.  Disinformation enters the model by allowing the individuals to  behave deceitfully,  which amounts to displaying a corrupted version of their environment's estimates.  This  harmful behavior happens with probability $\gamma$.  Copying corrupted information reduces the  chances of the copyist surviving an environmental challenge by  a factor $1/(1-\eta/2)$ on average, where $\eta \in [0,1]$  is the cost of believing corrupted information.   

Although the setup of  our individual-based model is presented in a manner  to stress its computer implementation, here  we offer an analytical study of the model, instead.  In particular, we derive an exact recursion equation  for the probability of survival of the  individuals in the limit of infinite population size and focus on  the time-asymptotic (equilibrium) solutions.  These solutions are characterized by the values of the optimal copying strategy $\tilde{w}$.   

Our main finding is that  the exploration-only pure strategy $\tilde{w}=0$ is optimal in a large region of the  space of model parameters, viz., in the regions where the probability of   behaving  deceitfully and the cost of believing corrupted information are moderate or large. Hence, by eroding trust,  disinformation  suppresses the copying or imitative behavior, which is vital for the success of our species\cite{Blackmore_2000} and, less importantly,  for  the design of efficient problem-solving heuristics.\cite{Fontanari_2014} This point is  neatly  expressed in the  phrase\cite{Bloom_2001}: ``Imitative learning acts like a synapse, allowing information to leap the gap from one creature to another''.  Contrasting these  ill-omened  results,  we find   that the other pure  strategy, viz., the copy-only strategy $\tilde{w}=1$ is  optimal when  the individuals are likely to behave honestly and the cost of copying corrupted information is high. This happens because the probability of survival of individuals in a totally credulous  population (i.e., $\tilde{w}=1$) does not depend on the cost $\eta$.  On the other hand, the population following a mixed strategy $0 < \tilde{w}<1$  is severely affected by the increase of  $\eta$. In fact, it is so badly affected that for large $\eta$ only the  pure strategies are  optimal, in which case increase of $\gamma$ leads to a discontinuous transition between the  copy-only and the exploration-only  strategies.

The rest of this paper is organized as follows. In  Sec.~\ref{sec:model}, we present the rules that govern the interaction of the individuals with their environment, as well as  with each other (e.g., through the copying and deceitful behaviors). We also  introduce  the  game evolutionary dynamics that determines how the survivors of the environmental challenges will  contribute  to the composition of  future generations.   In Sec.~\ref{sec:Analytical}, we derive  a recursion equation for the mean number of individuals that survive the environmental challenges at  generation $t$, which we   denote by $\Lambda^{(t)}(w)$. This quantity  gives a measure of  the  fitness of a population  using  the copying strategy $w$ and its maximization with respect to $w$ defines the optimal  copying strategy $\tilde{w}$ mentioned before.   In Sec.~\ref{sec:res}, we present the solutions of the recursion equation for $\Lambda^{(t)}(w)$. We  summarize our findings by presenting a phase diagram  in the space of the model parameters that shows the regions where the pure and the mixed strategies are optimal. 
 Finally, in Sec.~\ref{sec:disc} we review our main results and present some concluding remarks. 
 
\section{The Model}\label{sec:model}

We consider  a population of $N$ individuals that seek to assess some property of their environment, which   is essential to their survival  in some sense. In the context of animal behavior,\cite{Dall_2005}  this property could be  the abundance of a key resource,\cite{Clark_1984}  the risk of predation in the group's surrounding,\cite{Vine_1971} or cues to migratory decision in seasonal migration.\cite{Fryxell_1988} 
In the context of epistemic groups,\cite{Kitcher_1993,Reijula_2019,Fontanari_2019} which is the focus of the present contribution, that property could be the truthfulness of a  certain stance, e.g., the efficacy of vaccines or face masks to prevent transmission of the coronavirus. 

Our individual-based model builds on the following assumptions:

\begin{itemize} 

\item Each individual $i=1, \ldots,N$ gathers information about  the environment  either through direct exploration, i.e., trial and error learning,\cite{Thorndike_1898}  or through the observation and copy  of other individuals, i.e., social learning.\cite{Boyd_1985,Feldman_1996} Accordingly,  we denote  the probability that a target  individual copies another randomly chosen individual  (the model)   by $w \in [0,1]$. Hence,  $1-w$ is the probability that the target individual  explores the environment by itself.  

\item  The clues about the key property of the environment are produced by a normal distribution of  mean $\mu$ and variance $\sigma^2$. The true value of that property is the mean $\mu$,  which the individuals can estimate by sampling (i.e., exploring) the environment.\cite{Aguinaga_2021}  The closeness of the estimate $\xi \sim N( \mu, \sigma)$  to the true value $\mu$ determines if  an  individual  will  or will not survive the environmental challenge. Examples of  environmental challenge are encountering a predator or being exposed to the  coronavirus. 

\item If individual  $i$ samples the value $\xi_i \sim N( \mu, \sigma) $, then the probability  $S_i$ that it will survive the  environmental challenge  is given by 
\begin{equation}\label{S}
S_i = \exp \left [ -\frac{1}{2} \left ( \xi_i - \mu \right )^2 \right ]. 
\end{equation}
We refer to the random variable $S_i$ as the viability of individual $i$.\cite{Burger_1995}  It is straightforward to derive an explicit expression for the probability distribution of $S$, viz.,
\begin{equation}\label{PS}
P (S) = \frac{1}{\sqrt{\pi} \sigma^2}   \frac{S^{1/\sigma^2 -1}}{\sqrt{- \ln S^{1/\sigma^2}}}
\end{equation}
for $S \in [0,1]$.  The moments of $P(S)$, 
\begin{equation}\label{MomS}
\mathbb{E}_S(S^n)  = \int_0^1 S^n P(S) dS = \frac{1}{\sqrt{1 + n \sigma^2}} ,
\end{equation}
 will play a key role in the analytical study presented in Sec.~\ref{sec:Analytical}. 
We emphasize that since the distribution (\ref{PS}) does not depend on $\mu$, we can set $\mu =0$ without loss of generality.  We will refer to the variance $\sigma^2$ as the hazardousness of the environment, since the greater $\sigma^2$ is, the lesser the odds of an individual surviving the environmental challenge.

\item Aside from exploring the environment by producing a sample $\xi \sim N( \mu, \sigma) $ or, equivalently, by sampling a new viability $S$ with the distribution (\ref{PS}), the individuals can  face the environmental challenge by copying the estimate $\xi$ from another individual.  Thus, it is implicit   that the estimates of $\mu$ are publicly displayed by the individuals, which is a valid assumption in the epistemic context given the widespread use of social media to divulge personal viewpoints on practically any matter. 

\item    An  individual  displays a corrupted version of its  estimate of $\mu$ with probability $\gamma \in [0,1]$, which results in the decrease of the viability of the copyist by a (random) factor  $1/\epsilon$, where  $ \epsilon \sim \mbox{Uniform}(1-\eta, 1)$ with $\eta \in [0,1]$. In other words, by  copying an individual whose viability  is $S$, the copyist  has probability $\gamma$ of ending up with viability  $\epsilon S$ and probability $1-\gamma$ of  ending up with viability $S$. Hence the parameter  $\gamma$ measures the degree of deceitfulness  of the individuals,  whereas  $\eta$  measures the (detrimental) effect on viability of taking  distorted information at face value (i.e., the cost of believing  distorted information). We note that the maximum cost ($\eta=1$)  reduces the   viability of the copyist by a factor $2$ on average. The moments of the random variable $\epsilon$, 
\begin{equation}\label{Mome}
\mathbb{E}_\epsilon(\epsilon^n)  = \int_{1-\eta}^1  \epsilon^n  d \epsilon =  \frac{1}{(n+1)\eta} \left [ 1 - (1-\eta)^{n+1} \right ] ,
\end{equation}
will also be useful  in the analytical study of Sec.~ \ref{sec:Analytical}. 

\end{itemize}

Our  aim is to  investigate  whether there is an optimal information acquisition strategy  $w$ that maximizes the chances of an individual surviving the environmental challenge    in the hazardous and socially unreliable scenario introduced before.  To do that   we use   an evolutionary game  approach where the survivors contribute offspring to the next generation, who  are then subjected to  new challenges.\cite{Maynard_1982} In particular, survival to the environmental challenge is dictated by the individuals' viabilities $S$ as described before, but the survivors have equal probability to contribute  offspring to the next generation in the repopulating process that resets the population size to the fixed value  $N$.   We note that $w=0$ (exploration only) and $w=1$ (copying only) are pure strategies, whereas any value of $w$ in between those extremes represents a mixed strategy.\cite{Neumann_1944}

The game evolutionary dynamics proceeds as follows:

\begin{itemize}

 \item  At generation $t=0$,  each individual is assigned a random viability $S_i, i=1, \ldots, N$  according to the probability distribution  (\ref{PS}).  
 
 \item  Each individual decides independently whether to explore the environment or to copy another individual.  These  two alternatives  occur with probability $1-w$ and $w$, respectively.  
In the case individual $k$  decides to explore the environment, it  generates a new sample of the viability $S_k$ using the distribution (\ref{PS}), which then replaces its previous viability.
In the case  individual $k$  decides to copy,  it chooses  randomly one of the $N-1$ individuals  in the population, say, individual $l$. Then with probability $1-\gamma$  individual $l$ displays its uncorrupted estimate of $\mu$, so that the viability of individual $k$ becomes $S_k = S_l$, and with probability $\gamma$   individual $l$  displays a distorted version of its estimate of $\mu$, so that $S_k = \epsilon S_l$. The update of the viabilities is done simultaneously for all individuals.  

\item  Each individual is subjected to the environmental challenge that determines who  will have a chance to contribute offspring to generation $t=1$. Consider, for instance, individual $k$ whose viability is $S_k$ after the exploration/copying stage described in the previous item: it will survive the environmental challenge  at $t=0$  if $S_k > u_0$ where    $u_0 \sim \mbox{Uniform}(0, 1)$. As before, all individuals are tested simultaneously and so  at this point we can measure the fraction of individuals that survive the environmental challenge at generation $t=0$, which we denote by $\Lambda^{(0)}$. 

\item  We form generation $t=1$ by randomly selecting  $N$ individuals with replacement from the $N \Lambda^{(0)}$  survivors of the environmental challenge at $t=0$.  

\item Since we have now $N$ individuals characterized by the viabilities $S_i, i=1, \ldots, N$,  a situation similar to our starting point, we can repeat the  procedure above  to obtain  the population composition at generations  $t=2,3 \ldots$. 

\end{itemize}

Repetition of  the  game  evolutionary dynamics using independent samples of the  individuals' viabilities  in  the initial  population  allows us to obtain the mean fraction of individuals that survive the environmental challenge $\langle \Lambda^{(t)} \rangle $ at an arbitrary generation $t$.  This is the main quantity we focus on in this paper, since  it gives a measure of  how well   individuals using  copying strategy $w$ are adapted  to their environment. We recall that $w$ is a measure of the credulity of the individuals, $\gamma$  of their deceitfulness and $\eta$ of the cost of believing corrupted information.

Although we have described the model as an easy-to-implement  individual-based simulation for finite $N$,  in the next section we show how   it can be solved analytically in the limit $N \to \infty$.

\section{Analytical Solution for infinite population size}\label{sec:Analytical}

Since all individuals are equivalent in the statistical sense (i.e., they differ only by the assignment of the random viabilities at generation $t=0$, which are drawn from the same probability distribution),  we can equate the mean fraction of individuals that survive the environmental challenge  $\langle \Lambda^{(t)} \rangle $ with the probability that an arbitrary  individual  survives that challenge at generation $t$. It turns out  that this probability can be evaluated exactly in the limit of infinite population size $N \to \infty$,  as  described next.

We begin by calculating the probability   that an arbitrary  individual $k$  survives the  environmental challenge at generation $t=0$.   There are three possible events.  First,  individual $k$  explores the environment by producing a new  sample of its estimate of $\mu$ and, consequently, of the viability  $S_k$, so the probability  it survives is 
\begin{equation}
(1-w) P (S_k > u_0) = (1-w) S_k
\end{equation}
since   $u_0 \sim \mbox{Uniform}(0, 1)$. Second, individual $k$  copies a randomly chosen individual, say individual $l$,  who,  in turn, chooses to exhibit its true estimate of $\mu$. 
In this event, the probability of survival of individual $k$ is 
\begin{equation}
\frac{w}{N-1} (1-\gamma) P (S_l > u_0) = \frac{w}{N-1}(1-\gamma)  S_l ,
\end{equation}
where we have used that the probability of selecting a particular individual  $l$ as a model is $1/(N-1)$.
 Third,   individual $l$  chooses to exhibit a corrupted version  of its estimate of $\mu$. Then  individual $k$ survives the challenge  with  probability 
\begin{equation}
\frac{w}{N-1} \gamma P (\epsilon_{l,0} S_l > u_0) = \frac{w}{N-1} \gamma  \epsilon_{l,0} S_l ,
\end{equation}
where $ \epsilon_{l,0} \sim \mbox{Uniform}(1-\eta, 1)$.
Putting the  contributions of these three events together, we can write down the (conditional) probability that individual $k$ survives  the environmental challenge given $S_i$ and $\epsilon_{i,0}$ for  $ i=1, \ldots, N$, viz.,
\begin{equation}
\Lambda_k^{(0)} = (1-w) S_k + \frac{w}{N-1} \sum_{l \neq k} S_l \left ( 1 - \gamma + \gamma \epsilon_{l,0} \right ) .
\end{equation}
Hence the (conditional)  mean  fraction of the population that survive the environmental challenge at $t=0$ is 
\begin{equation}\label{ll0}
\Lambda^{(0)} =  \frac{1}{N} \sum_{k=1}^N  \left [  1 \times \Lambda_k^{(0)}  + 0  \times (1  - \Lambda_k^{(0)}) \right ]  =  \frac{1}{N} \sum_{k=1}^N   \Lambda_k^{(0)}.
\end{equation}

At this point, it is important to note  that the probability that an individual that survived the environmental challenge at generation $t=0$  contributes an offspring to generation $t=1$ is  $1/ \left [ N  \Lambda^{(0)}  \right ]$, so that each survivor contributes  $1/  \Lambda^{(0)} $ offspring  to generation $t=1$, on the average. 

Finally, averaging Eq.~(\ref{ll0}) over  the statistically independent random variables $S_i$ and  $\epsilon_{i,0}$  and taking the limit $N \to \infty$  yield the unconditional  mean  fraction of the population that survive the environmental challenge at $t=0$,
\begin{equation}\label{L1}
\langle \Lambda^{(0)}  (w) \rangle  =  (1-w) \mathbb{E}_S(S) + w \mathbb{E}_S(S) \left [ 1 - \gamma + \gamma \mathbb{E}_\epsilon (\epsilon) \right ].
\end{equation}
Here the moments of the random variables $S$ and $\epsilon$ are given in Eqs.~(\ref{MomS}) and (\ref{Mome}), respectively.

The next step is to calculate the probability  that  target individual $k$ survives the  environmental challenge at generation $t=1$. The simplest possibility is that  individual $k$ explores the environment, so that its  probability of surviving the challenge at $t=1$  is simply  $(1-w)  S_k$, with $S_k$ distributed according to the distribution  (\ref{PS}).  
The alternative is  that individual $k$ copies some individual $l$ (the model), who may or may not exhibit its true estimate of $\mu$. Clearly, the probability of choosing a model  with viability $S_l$ equals the probability of choosing an offspring of the survivor with  viability $S_l$, viz.,  $1/ \left [ (N-1)  \Lambda^{(0)}  \right ]$. Assuming that the survivor has  explored the environment in the previous generation ($t=0$, in this case),  the probability  that the copyist survives the challenge at $t=1$ is 
\begin{eqnarray}\label{A21}
&  &   \frac{w (1-w)}{ (N-1) \Lambda^{(0)}} \sum_{l \neq k} P(S_l > u_0)  \left [  ( 1- \gamma) P(S_l > u_1)   + \gamma  P( \epsilon_{l,1} S_l > u_1 ) \right ] \nonumber \\
& & =  \frac{w (1-w)}{ (N-1) \Lambda^{(0)}}  \sum_{l \neq k} S_l^2 \left [  1- \gamma    + \gamma  \epsilon_{l,1}  \right ] ,
\end{eqnarray}
where the index of the uniform random variable $u$   determines the generation at which the challenge was taken. Similarly, the second index of the random variables  $\epsilon$ determines the generation at which a corrupted estimate of $\mu$  was  produced.
In sum, Eq.~(\ref{A21}) yields the probability that target individual $k$ survives the challenge by copying the offspring of an individual that has explored the environment in the previous generation.  Let us assume now that  the ancestor of model $l$ has copied an authentic estimate of $\mu$  from  another individual $m$ in the previous generation.  In this case, the probability that target individual $k$ survives the challenge at $t=1$ is
\begin{eqnarray}\label{A22}
&  & \frac{w^2 (1- \gamma) }{(N-1)(N-1) \Lambda^{(0)}}   \sum_{l \neq k}  \sum_{m \neq l} P(S_m > u_0) 
 \left [  ( 1- \gamma) P(S_m > u_1)  \right. \nonumber \\
 &  &   \hspace{6.2cm} \left. + \gamma  P( \epsilon_{m,1} S_m > u_1 ) \right ] 
  \nonumber \\
& & =  \frac{w^2 (1- \gamma) }{(N-1)(N-1) \Lambda^{(0)}}   \sum_{l \neq k}  \sum_{m \neq l} S_m^2 \left [  1- \gamma    + \gamma  \epsilon_{m,1}  \right ] .
\end{eqnarray}
We note that in this equation individuals $k$ and $l$ belong to generation $t=1$,  whereas individual $m$ belongs to generation $t=0$. Equation (\ref{A22}) yields the probability that individual $k$ survives the challenge at $t=1$ by copying the offspring of an individual that has copied an authentic estimate of  $\mu$ from another individual  in generation $t=0$.
Lastly, we must consider the possibility that  the ancestor of model $l$ has copied a corrupted estimate of $\mu$  from  another individual $m$ in the previous generation. In this case the probability that individual $k$ survives the challenge at $t=1$ is
\begin{eqnarray}\label{A23}
&  & \frac{w^2 \gamma }{(N-1)(N-1)  \Lambda^{(0)}}   \sum_{l \neq k}  \sum_{m \neq l} P(\epsilon_ {m,0} S_m > u_0) 
 \left [  ( 1- \gamma) P(\epsilon_{m,0} S_m > u_1)   \right. \nonumber \\
 &  &   \hspace{7cm} \left. + \gamma  P( \epsilon_{l,1} \epsilon_{m,0} S_m > u_1 ) \right ] 
  \nonumber \\
& & =  \frac{w^2 \gamma }{(N-1)(N-1) \Lambda^{(0)}}   \sum_{l \neq k}  \sum_{m \neq l} \epsilon_{m,0}^2 S_m^2 \left [  1- \gamma    + \gamma  \epsilon_{l,1}  \right ] .
\end{eqnarray}
As before, this equation  yields the probability that individual $k$ survives the challenge at $t=1$ by copying the offspring of an individual that has copied a corrupted estimate of  $\mu$ from another individual  in generation $t=0$.

The obstacle to proceed further, i.e., to carry out  the averages over   the statistically independent random variables $S_i$,  $\epsilon_{i,0}$  and $\epsilon_{i,1}$ is the presence of the term $\Lambda^{(0)}$ in the denominator of Eqs.~(\ref{A21}),  (\ref{A22}) and (\ref{A23}).
To circumvent  this difficulty, we will assume that the random variable    
\begin{equation}
\Lambda^{(t)}  \equiv  \frac{1}{N} \sum_{k=1}^N   \Lambda_k^{(t)} 
\end{equation}
is self-averaging in the limit $N \to \infty$, i.e., $\Lambda^{(t)}  \to \langle \Lambda^{(t)} \rangle $ for all $t$. The evidence we offer to support this assumption is the excellent agreement  between our analytical predictions and the individual-based simulations for large $N$ (see, e.g., Figs. \ref{fig:1} and \ref{fig:2}). However, for $t=0$, at least,  the self-averaging assumption follows directly from the law of large numbers.\cite{Feller_1968} 

Finally, averaging  over $S_l$ and $\epsilon_{i,\tau}$ and assuming that $N$ is large so that we can write $ \Lambda^{(0)} \approx  \langle \Lambda^{(0)}\rangle $ we obtain the probability that a randomly chosen individual survives the environmental challenge at generation $t=1$,  
\begin{equation}\label{L2}
\langle \Lambda^{(1)}  (w) \rangle   =   (1-w) \left [  \mathbb{E}_S (S) +  \frac{w}{\langle \Lambda^{(0)}\rangle}  b_1 \mathbb{E}_S (S^2) \right ] 
+  \frac{w^2}{\langle \Lambda^{(0)}\rangle} b_1 b_2 \mathbb{E}_S (S^2) 
\end{equation}
where 
\begin{equation}\label{bi}
b_\tau = 1- \gamma + \gamma \mathbb{E}_\epsilon (\epsilon^\tau) .
\end{equation}
The same reasoning leads to the probability that an individual survives the environmental challenge at generation $t=2$,
\begin{eqnarray}\label{L3}
\langle \Lambda^{(2)}  (w) \rangle   &  = &    (1-w) \left [  \mathbb{E}_S (S)  +  \frac{w}{\langle \Lambda^{(1)}\rangle}  b_1 \mathbb{E}_S (S^2)
+  \frac{w^2}{\langle \Lambda^{(1)}\rangle \langle \Lambda^{(0)} \rangle } b_1 b_2 \mathbb{E}_S (S^3) \right ] \nonumber \\
&  &   \mbox{} + \frac{w^3}{\langle \Lambda^{(1)}\rangle \langle \Lambda^{(0)} \rangle } b_1 b_2 b_ 3 \mathbb{E}_S (S^3),
\end{eqnarray}
and at generation $t=3$,
\begin{eqnarray}\label{L4}
\langle \Lambda^{(3)}  (w) \rangle   &  = &    (1-w) \left [  \mathbb{E}_S (S)  +  \frac{w}{\langle \Lambda^{(2)}\rangle}  b_1 \mathbb{E}_S (S^2)
+  \frac{w^2}{\langle \Lambda^{(2)}\rangle \langle \Lambda^{(1)} \rangle } b_1 b_2 \mathbb{E}_S (S^3)  \right . \nonumber \\
& &   \hspace{4cm} \left .
  + \frac{w^3}{\langle \Lambda^{(2)}\rangle \langle \Lambda^{(1)} \rangle \langle \Lambda^{(0)} \rangle} b_1 b_2 b_3 \mathbb{E}_S (S^4)
\right ] \nonumber \\
&  &   \mbox{} + \frac{w^4}{\langle \Lambda^{(2)}\rangle \langle \Lambda^{(1)} \rangle  \langle \Lambda^{(0)} \rangle } b_1 b_2 b_ 3 b_4 \mathbb{E}_S (S^4) .
\end{eqnarray}
The idea behind the derivation of these expressions is to follow  the ancestry of the target  individual living at generation $t$  until we find an ancestor that  explored the environment. Here, ancestry is determined by  who copy whom.   Assume that this ancestor lived at generation  $0 \leq \tau \leq t$ and that $\gamma = 0$ (hence $b_i=1$), for simplicity.   Clearly, the probability of this happening is proportional to $w^{t-\tau} (1-w)$.  In addition, the viability $S$ drawn by the ancestor at generation $\tau$ must pass $t - \tau + 1$ independent challenges and the descendants  of the ancestor in the lineage of the target individual must  be chosen  in all the repopulation  stages that take place from $\tau$ to $t-1$. The probability that this happens is
\begin{equation}
\frac{P(S>u_{\tau})}{ \langle \Lambda^{(\tau)} \rangle} \times \frac{P(S>u_{\tau+1})}{ \langle \Lambda^{(\tau+1)} \rangle} \times \ldots
\times \frac{P(S>u_{t-1})}{ \langle \Lambda^{(t-1)} \rangle} \times P(S>u_{t})
\end{equation}
which reduces to $S^{t - \tau + 1}/ \left [  \langle \Lambda^{(\tau)} \rangle  \langle \Lambda^{(\tau+1)} \rangle  \ldots  \langle \Lambda^{(t-1)} \rangle \right ] $ since $u_i \sim \mbox{Uniform}(0, 1)$.    
Of course, there is also the possibility that all  ancestors of the target individual down to generation  $t=0$ are copyists, which happens with probability proportional to $w^{t+1}$, and is accounted for  by the last terms in Eqs. (\ref{L1}), (\ref{L2}), (\ref{L3}) and (\ref{L4}). At this stage, we can readily write down the generalization of those recursion equations for  arbitrary $t$,
\begin{equation}\label{rec}
\langle \Lambda^{(t)}  (w) \rangle   = (1-w) \sum_{\tau=0}^{t} a_{\tau,t} \mathbb{E}_S(S^{\tau+1}) w^\tau   + a_{t+1,t}  \mathbb{E}_S(S^{t+1})  w^{t+1}
\end{equation}
with $a_{0,t} = 1$,  
\begin{equation}
a_{\tau,t} = \frac{b_\tau}{ \langle \Lambda^{(t-\tau)} \rangle }  a_{\tau-1,t}
\end{equation}
for $\tau >1$,  and we have defined   $\langle \Lambda^{(-1)} \rangle = 1$.  In Sec. \ref{sec:res} we offer explicit expressions for $\langle \Lambda^{(t)} \rangle $ in the limiting cases $w=0$ and $w=1$, as well as the first terms of its expansion in powers of $w$.

\section{Results}\label{sec:res}

\begin{figure}[]
\centerline{\includegraphics[width=.6\textwidth]{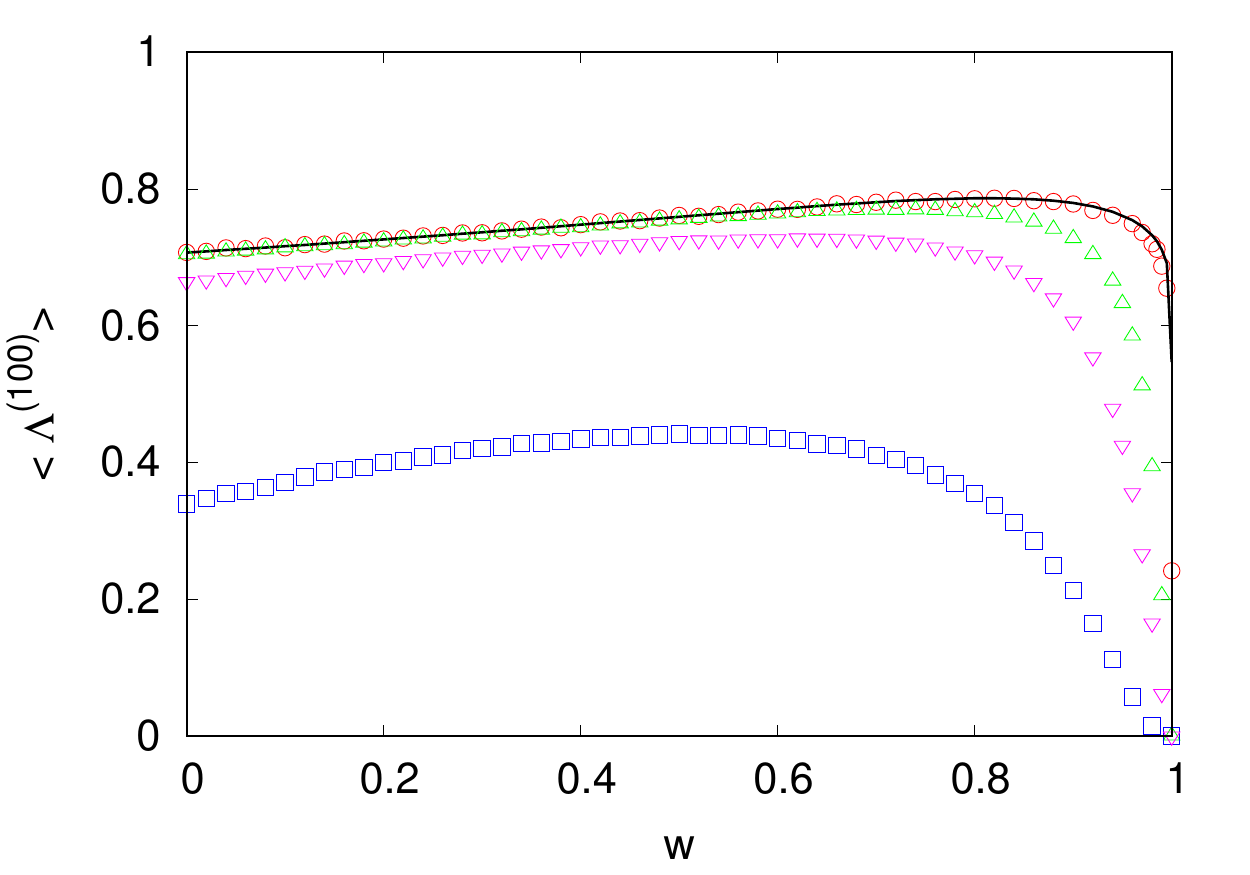}}
\vspace*{8pt}
\caption{Mean fraction  of individuals that survive the environmental challenge at generation $t=100$ against the copying probability $w$ for $N=4~ (\square)$, $N=6~ (\triangledown)$, $N=8 ~(\triangle)$ and $N=100 (\bigcirc)$. The solid curve is the analytic prediction for $N \to \infty$. The other parameters are $\gamma =0.5$, $\eta = 0.1$ and $\sigma^2 = 1$.}
\label{fig:1}
\end{figure}

Figure \ref{fig:1} shows the effect of the population size $N$ on the   mean fraction  of individuals that survive the environment challenge at generation $t=100$. As pointed out, $\langle \Lambda^{(t)}\rangle$  is a proxy for the quality of the adaptation of the individuals to their environment, whereas the copying probability $w$ is a proxy for the credulity of the individuals.  In this figure, we set the deceitfulness of the individuals to $\gamma = 0.5 $ and the cost of copying (or believing)   corrupted information to $\eta = 0.1$. This means that the individuals have a $50\%$ chance of behaving deceitfully and that the survival probability of a copyist is reduced by a factor   of  $1/(1-\eta/2) = 100/95 \approx 1.05$ on the average when copying a deceitful  model. 

\begin{figure}[]
\centerline{\includegraphics[width=.6\textwidth]{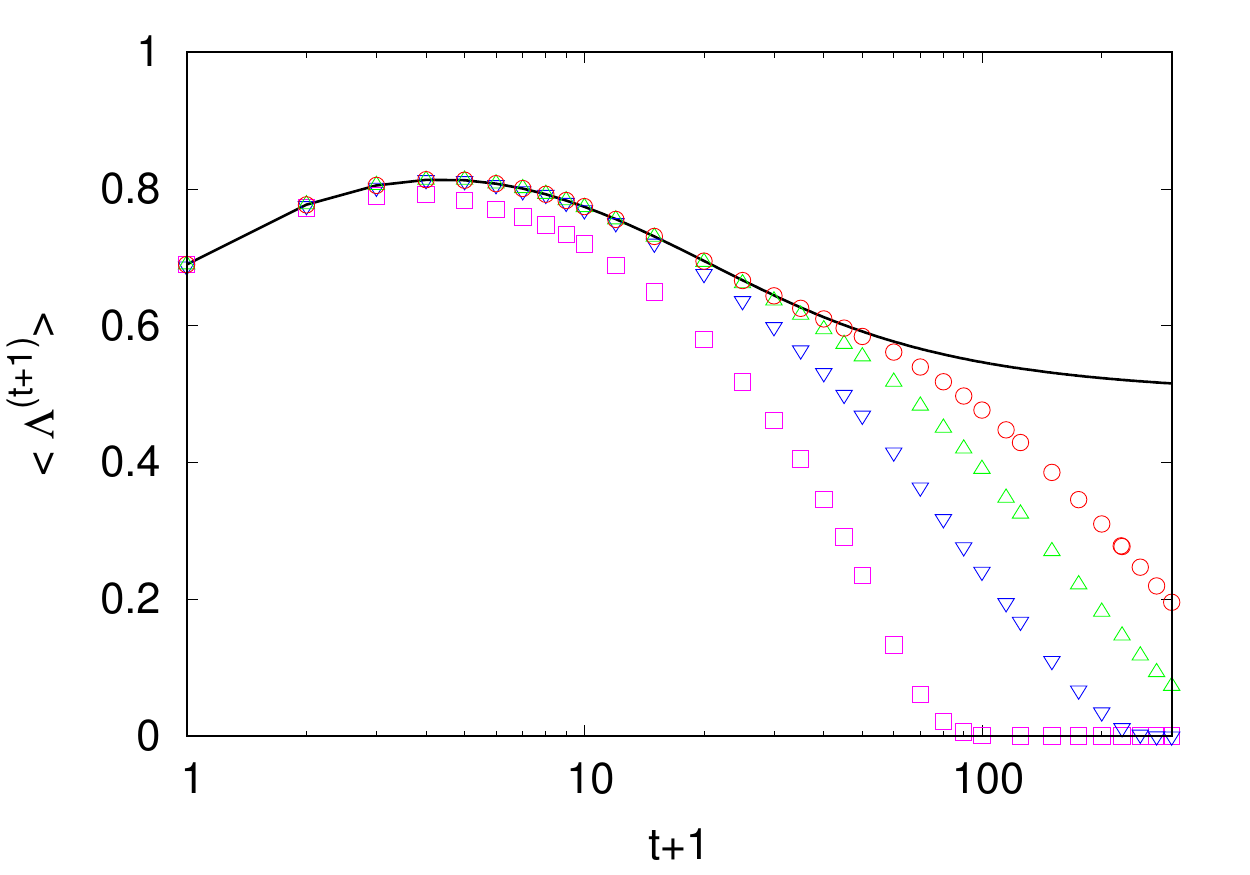}}
\vspace*{8pt}
\caption{Mean fraction  of individuals that survive the environmental challenge as function of the  generation $t \leq 300$ for a population of totally credulous individuals ($w=1$) of size  $N=10~ (\square)$, $N=10^2~ (\triangledown)$, $N=10^3 ~(\triangle)$ and $N=10^4 (\bigcirc)$. The solid curve is the analytic prediction (\ref{w1}) for $N \to \infty$. The other parameters are $\gamma =0.5$, $\eta = 0.1$ and $\sigma^2 = 1$. In the limit $t \to \infty$  we find  $\langle\Lambda^{(t)} (1) \rangle  \to 0.5$. The generation index $t$ was shifted  by one unit to the right to allow use of  the logscale in the x-axis.  }
\label{fig:2}
\end{figure}

The primary purpose of Fig. \ref{fig:1}  is to show the excellence of the analytical prediction for large $N$:  already for $N=100$ the individual-based simulations depart significantly from that prediction only in the region very close to  $w =1$. We note that  for the small population sizes  considered (viz.,  $N=4, 6$ and $8$),  all runs resulted in extinction for $w=1$, whereas  no extinction was observed for $N=100$. This is expected since for $w=1$  the individuals sample their environment at the initial generation $t=0$ only  and so their viabilities $S_k, k=1, \ldots,N$  can never be greater than the maximum of $N$ samples drawn from the  distribution (\ref{PS}). Of course, for sufficiently large $t$, the runs with $N=100$ (and $w=1$) should all result in extinction as well (see Fig.\ \ref{fig:2}).  This is not the case for $N \to \infty$, however. In fact, in this limit   we can write an explicit equation for $\langle \Lambda^{(t)} (1) \rangle $, viz.,
\begin{eqnarray}\label{w1}
\langle \Lambda^{(t)} (1) \rangle  & = & \left [ 1 - \gamma + \gamma \mathbb{E}_\epsilon (\epsilon^{t+1}) \right ] \frac{ \mathbb{E}_S (S^{t+1}) }{ \mathbb{E}_S (S^{t}) }
\nonumber \\
& = & \left [ 1 - \gamma + \frac{\gamma}{(t+2)\eta} [ 1 - (1-\eta)^{t+2}] \right ] \sqrt{ \frac{1 + t \sigma^2}{1 + (t+1) \sigma^2}},
\end{eqnarray}
from where we obtain  
\begin{equation}\label{w11}
\lim_{t \to \infty} \langle \Lambda^{(t)} (1) \rangle   = 1 - \gamma ,
\end{equation}
 provided that $\eta >0$. (For  $\eta =0$, Eq.~(\ref{w1}) yields $\lim_{t \to \infty} \lim_{\eta \to 0} \langle \Lambda^{(t)} (1) \rangle   = 1$.)  Thus, extinction is not certain if $\gamma < 1$, i.e., if there is a nonzero chance that the individuals will behave honestly. Nevertheless, a low value of $\langle \Lambda^{(t)} (1) \rangle $ points to the maladaptation of a population of  totally credulous individuals. Figure \ref{fig:2}  summarizes these results. We stress that the  observed  slow convergence to the infinite population size limit  as $N$ increases   happens only for $w=1$, because a finite population is certain to go extinct in this case. For $w < 1$, the convergence to the infinite population size limit is much faster, as illustrated in Fig.~\ref{fig:1}.

Another limiting case for which we can obtain an explicit solution of the recursion equation (\ref{rec}) is $w=0$, which  describes  the situation  where the individuals never copy their peers. In this case  we have
\begin{equation}\label{w0}
\langle \Lambda^{(t)} (0)  \rangle =  \mathbb{E}_S (S) = \frac{1}{\sqrt{1 + \sigma^2}},
\end{equation}
which does not depend on the generation $t$, as expected, since at each generation all individuals produce new estimates of $\mu$, as in the setup of the  initial generation. It is interesting to note that in our model the popular notion of culture  as  information and experiences transfered from the older generations to the newer generation\cite{Kroeber_1952,Blackmore_2000} is intimately associated to a nonzero value of the  copying probability $w$, which means that  culture   requires individuals that possess some degree of credulity.

Henceforth we will consider only the limit of infinite population size $N \to \infty$,  for which  the recursion equation (\ref{rec}) holds true and can easily be iterated numerically to produce the  mean fraction  of individuals that survive the environmental challenge at any  generation $t$.  In  particular, we will focus on the  equilibrium  regime, which we can  assume  to be reached at generation  $t=1000$ (see Fig.~\ref{fig:2}), so we can write $\langle \Lambda^{(\infty)}(w)  \rangle  \approx \langle \Lambda^{(1000)} (w) \rangle $ for all $w \in [0,1]$. Although this is a safe assumption in general, it fails to produce high precision estimates of $\langle \Lambda^{(\infty)} (w) \rangle $ for $w \approx 1$ that are needed to determine the values of some critical parameters, as we will see later in this section. In that case we assume that the equilibrium  is reached for  $t=10^4$.

\begin{figure}[]
\centerline{\includegraphics[width=.6\textwidth]{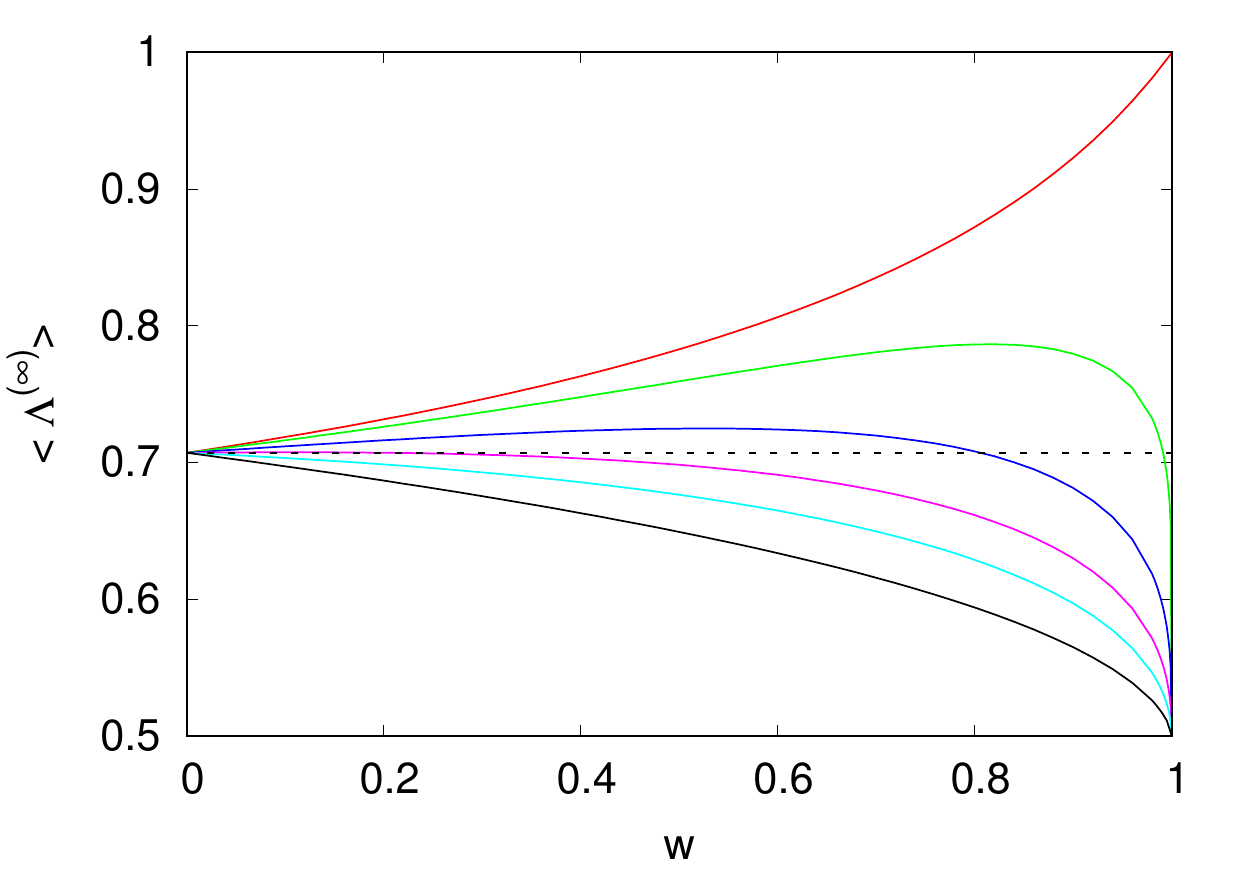}}
\vspace*{8pt}
\caption{Mean fraction of individuals that survive the environmental challenge  for a population of infinite size at equilibrium against 
the copying probability $w$ for $\gamma =0.5$, $\sigma^2 = 1$ and (top to bottom) $\eta =0, 0.1, 0.3, 0.5, 0.7$ and $1$.
 The dashed horizontal  line is  $ \langle \Lambda^{(\infty)} (0) \rangle=  1/\sqrt{2} \approx 0.707$.  
 For  $\eta > 0$ we have   $ \langle \Lambda^{(\infty)}(1) \rangle  = 0.5$, so that   $\langle \Lambda^{(\infty)}(0)  \rangle >  \langle \Lambda^{(\infty)} (1) \rangle $. The critical cost of copying corrupted information beyond which  epistemic security is lost  is $\eta_c^0 = 4-2\sqrt{3} \approx  0.536$. }
\label{fig:3}
\end{figure}

Figure \ref{fig:3} shows the analytical predictions for the fraction of individuals that survive the environmental challenge at equilibrium for  $\gamma = 0.5$, $\sigma^2 =1$    and different values of the cost $\eta$ of believing distorted information. These results  reveal  that, given the parameters $\gamma$ and $\eta$, there is an optimal copying  probability $\tilde{w}$ that maximizes the adaptation of the population to the environment. In the case $\eta=0$,  which describes the situation where there is no cost to copying corrupted information (e.g., because it is irrelevant for survival),  the probability of survival is maximized by a population of totally credulous individuals, i.e., $\tilde{w}=1$. 
As $\eta$ increases, it pays for the individuals to be  more skeptical  so that $\tilde{w} < 1$. The remarkable result here  is that above a critical  cost $\eta_c^0$ we find  $\tilde{w}=0$. In other words, the optimal strategy for survival is not copying at all, which means that culture (or an epistemic society) would never emerge since, as pointed out before,  culture requires the copying of  information produced by older generations.\cite{Kroeber_1952,Blackmore_2000}

\begin{figure}[]
\centerline{\includegraphics[width=.6\textwidth]{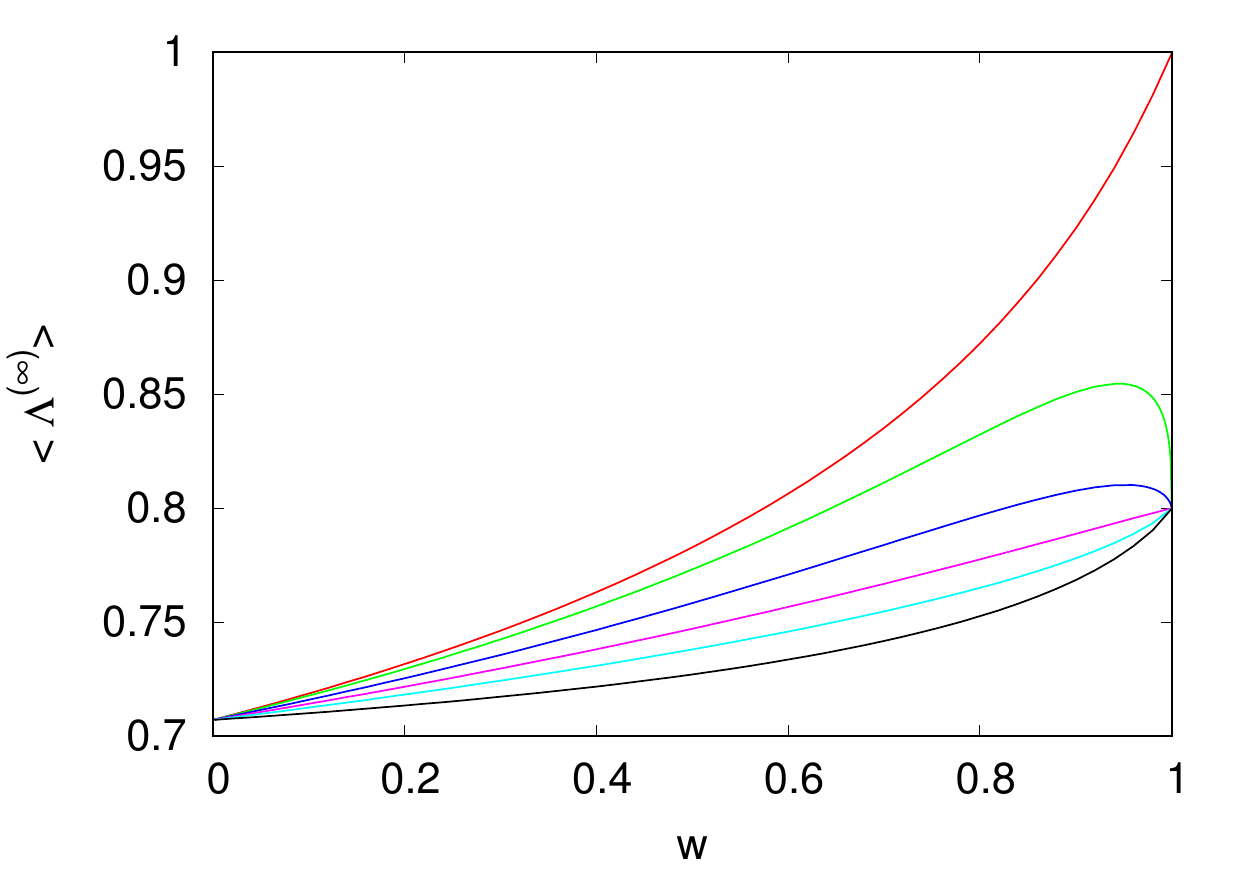}}
\vspace*{8pt}
\caption{Mean fraction of individuals that survive the environmental challenge  for a population of infinite size at equilibrium against 
the copying probability $w$ for $\gamma =0.2$, $\sigma^2 = 1$ and (top to bottom) $\eta =0, 0.1, 0.3, 0.5, 0.7$ and $1$.  Here $ \langle \Lambda^{(\infty)} (0) \rangle =  1/\sqrt{2} < \langle \Lambda^{(\infty)} (1) \rangle  = 0.8$.
 The critical cost of copying corrupted information beyond which survival is maximized by totally credulous individuals is  $\eta_c^1  \approx  0.480$. }
\label{fig:4}
\end{figure}

\begin{figure}[]
\centerline{\includegraphics[width=.6\textwidth]{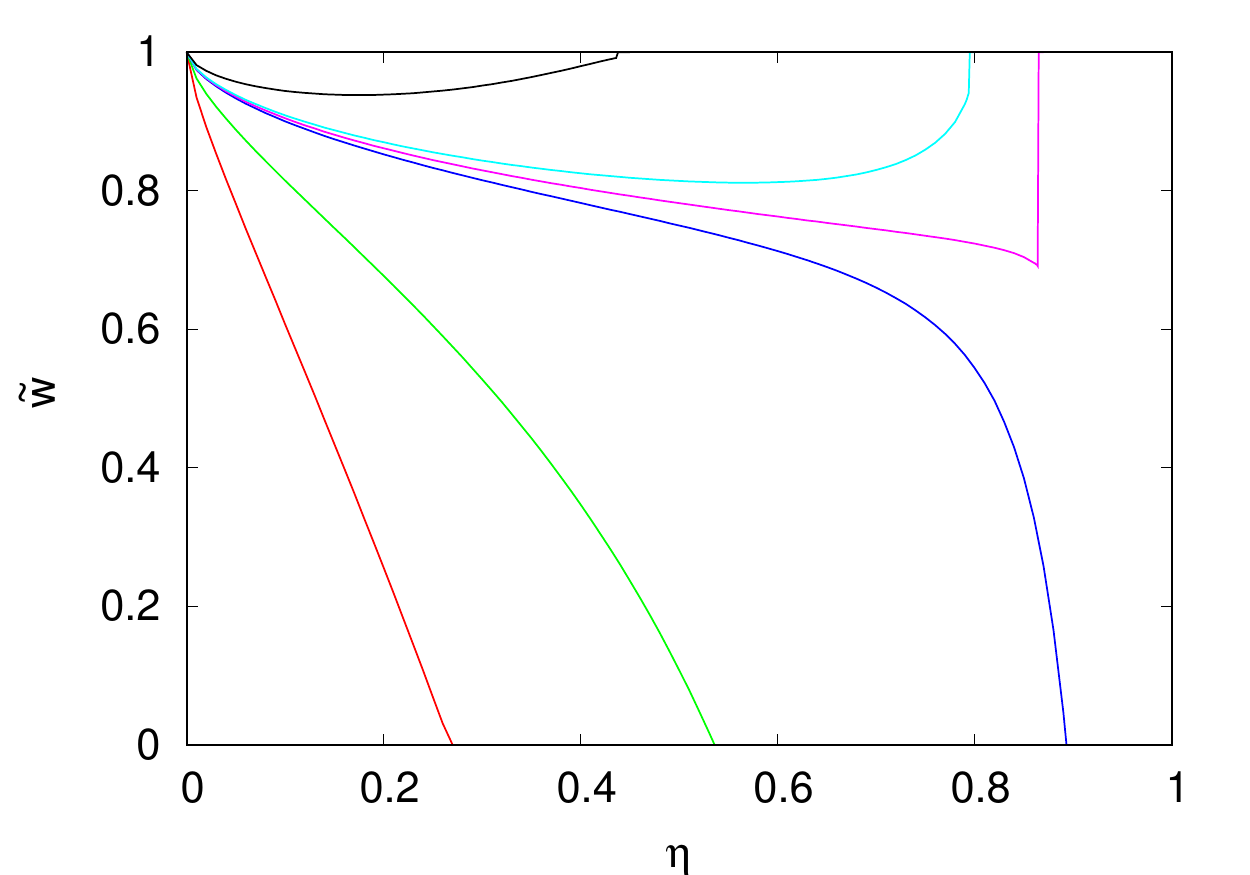}}
\vspace*{8pt}
\caption{Optimal copying probability $\tilde{w}$  against the cost of copying corrupted information  $\eta$ for $\sigma^2 = 1$ and (top to bottom)
$\gamma  = 0.2, 0.28, 0.29, 0.3, 0.5$ and $1$. For $ \gamma >  1-1/\sqrt{2} \approx 0.293$ there is a continuous transition at $\eta_c^0$ to a regime where epistemic security is lost ($\tilde{w}=0$),  whereas for $ \gamma <  1-1/\sqrt{2}$  there is a transition at $\eta_c^1$ to a regime where the individuals are totally credulous ($\tilde{w}=1$). This transition is continuous for $\gamma=0.2$ and discontinuous for $\gamma=0.28$ and $0.29$.
  }
\label{fig:5}
\end{figure}

However, the scenario depicted in Fig.~\ref{fig:3}  happens only  for  $ \langle \Lambda^{(\infty)} (1) \rangle    < \langle \Lambda^{(\infty)} (0)  \rangle  $ that corresponds to the condition (see Eqs.~(\ref{w11}) and (\ref{w0}))
\begin{equation}\label{ine1}
  \gamma  > 1 - \frac{1}{\sqrt{1+\sigma^2}} ,
\end{equation}
which is satisfied when the probability of behaving deceitfully is high and the hazardousness of the environment is low.
The scenario where  condition (\ref{ine1}) is violated is illustrated in Fig.~\ref{fig:4}, which shows  $\langle \Lambda^{(\infty)} \rangle$ for $\gamma = 0.2$   and the same values of $\sigma^2$  and  $\eta$ used in Fig.~\ref{fig:3}. As  the cost $\eta$  increases from $0$ to $1$, the optimal copying probability $\tilde{w}$ initially decreases indicating the advantage of some degree of skepticism, as expected, but  then it starts to increase  and at a critical value $\eta_c^1$ it reaches  $1$,  remaining fixed at  that value
as $\eta$ increases further towards $1$  (see curve for $\gamma =0.2$ in Fig.~\ref{fig:5}).  This  apparently counter-intuitive result has a simple explanation: since $\langle \Lambda^{(\infty)} (1) \rangle $ does not depend on the cost $\eta$ (see Eq.~(\ref{w11})),  increase of  $\eta$ does not affect the survival of a population of  totally credulous  individuals ($w=1$), while it is detrimental to a population of skeptical individuals ($w < 1$), as shown in Fig.~\ref{fig:4}. To understand why   $\langle \Lambda^{(\infty)} (1) \rangle $ does not depend on $\eta$, we note  that for $\gamma < 1$  there is always a chance that at least one individual behaves honestly, in which case the copied information is authentic. The fortunate copyist  is very likely to get through the environmental challenge and, in the worst-case scenario that  all the other individuals fail that challenge, repopulate the entire population with its clones.

Figure \ref{fig:5} illustrates the complex influence of the parameters $\gamma$ and $\eta$ on the optimal copying strategy $\tilde{w}$. As pointed out before, there are two very distinct scenarios depending on whether  $\langle \Lambda^{(\infty)}(0)  \rangle $  is less or greater than $ \langle \Lambda^{(\infty)} (1)\rangle $.  In fact, given that $\tilde{w}$ is obtained by maximizing the mean fraction of survivors $\langle \Lambda^{(\infty)} (w)  \rangle $, it is clear that $\tilde{w} < 1 $ if $\langle \Lambda^{(\infty)} (0)  \rangle > \langle \Lambda^{(\infty)} (1) \rangle $ (the trivial exception is for $\eta=0$) and  $\tilde{w} >0$, otherwise. In the first scenario, where  the probability of behaving deceitfully is high,  increasing the cost  of copying corrupted information  gives rise to  a continuous phase transition to a regime characterized by individuals that solely explore their environment,  because copying is too risky.  This  doomsday scenario where culture is lost is characterized by the pure strategy $w = 0$. Surprisingly, in the second scenario, where the individuals are likely to behave honestly,  increasing  the cost $\eta$  prompts a transition to a regime where the population is composed of totally credulous individuals, i.e.,  the pure strategy $w = 1$ is optimal. The transition is continuous for small $\gamma$  but becomes discontinuous with increasing $\gamma$.  In particular,  for large $\eta$, say $\eta =1$, $\tilde{w}$ jumps from $0$ to $1$ at  $\gamma = 1-1/\sqrt{1+\sigma^2}$.

\begin{figure}[]
\centerline{\includegraphics[width=.6\textwidth]{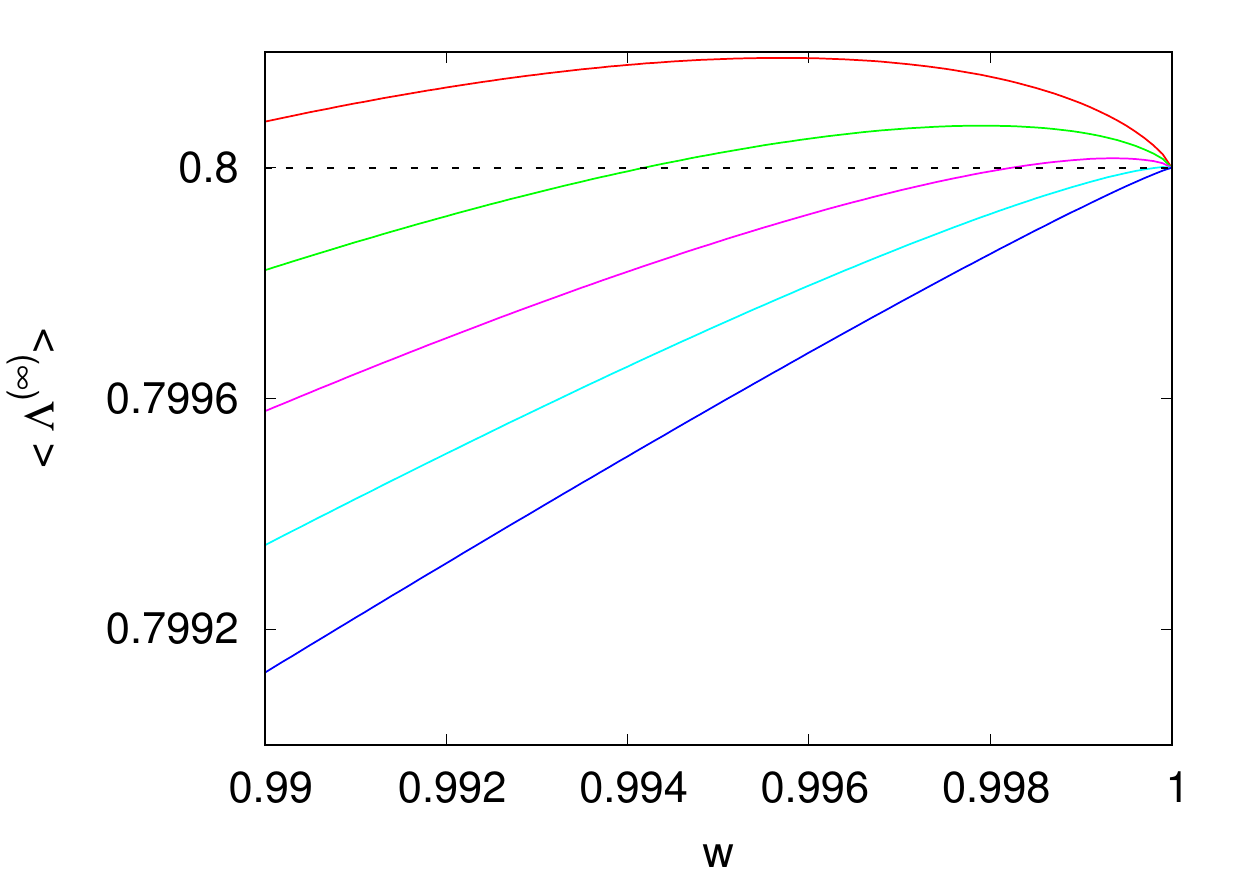}}
\vspace*{8pt}
\caption{Mean fraction of individuals that survive the environmental challenge  for a population of infinite size at equilibrium against 
the copying probability $w$ for $\gamma =0.2$, $\sigma^2 = 1$ and (top to bottom) $\eta =045, 0.46, 0.47, 0.48 $ and $0.49$.
 The dashed horizontal  line is  $ \langle \Lambda^{(\infty)}(1) \rangle =  1-\gamma  =0.8$.  The point at which the continuous  transition takes place is $\eta_c^1 \approx 0.480$.
  }
\label{fig:6}
\end{figure}

In the case that condition (\ref{ine1}) is satisfied, i.e., for $ \gamma >  1-1/\sqrt{1+\sigma^2}$, we can derive analytical expressions for the critical point $\eta_c^0$, as well as for the optimal copying probability $\tilde{w}$ near the critical point. In fact,  examination of Fig.~\ref{fig:3} reveals that the condition for $\tilde{w}>0$ is that the derivative of $\langle \Lambda^{(\infty)}(w) \rangle$  calculated at $w=0$ be positive, so that  the critical cost $\eta_c^0$ can be determined by setting  that derivative to zero. 
To take advantage of this observation, we need to expand $\langle \Lambda^{(t)} \rangle$ in powers of $w$.   It is easy to see that to obtain the correct terms of  order  $w^{k}$ with $k=0, \ldots, t$,  it is enough to expand  $\langle \Lambda^{(k)} \rangle$ to order $w^{k}$. In other words, the coefficients of  the powers $w^0$, $w^1$ and $w^2$  are the same in the expansion of 
 $\langle \Lambda^{(2)} \rangle$  as in the expansion of $ \langle \Lambda^{(\infty)}\rangle$. We find
\begin{equation}\label{expan}
\langle \Lambda^{(\infty)} \rangle  \approx  \mathbb{E}_S(S) - w  \mathbb{E}_S(S) \left [ 1 - b_1 \frac{\mathbb{E}_S(S^2)}{\mathbb{E}_S^2(S)} \right ] 
-w^2 b_1 \left [ b_1 \frac{\mathbb{E}^2_S(S^2)}{\mathbb{E}^3_S(S)} - b_2 \frac{\mathbb{E}_S(S^3)}{\mathbb{E}^2_S(S)} \right ]
\end{equation}
with $b_i$ defined in Eq.~(\ref{bi}).
 Hence  the critical or threshold  parameter is determined by the equation  
\begin{equation}\label{b1c}
b_1 = \frac{\mathbb{E}_S^2(S)}{\mathbb{E}_S(S^2)} 
\end{equation}
 that can be rewritten as
\begin{equation}\label{etac}
\eta_c^0 = \frac{2}{\gamma} \left ( 1 - \frac{\sqrt{1+2\sigma^2}}{1 + \sigma^2} \right ) .
\end{equation}
We note that   condition $\eta_c^0 \geq 0$ is always satisfied,  but condition $\eta_c^0 \leq 1$ is satisfied  only for
\begin{equation}\label{ine2}
\gamma > 2  \left ( 1 - \frac{\sqrt{1+2\sigma^2}}{1 + \sigma^2} \right ) .
\end{equation}
Hence the maximum value that $\eta_c^0$ can take on is determined by the lower bound of $\gamma$, which is given either  by inequality (\ref{ine1})  or by  inequality (\ref{ine2}). Explicitly, for $\sigma^2 < 1.218$  inequality (\ref{ine1}) holds and the maximum of $\eta_c^0$ is 
\begin{equation}\label{eta_max}
\tilde{\eta}_c^0 = 2 \left ( 1 - \frac{\sqrt{1+2\sigma^2}}{1 + \sigma^2} \right ) \left ( 1 - \frac{1}{\sqrt{1+\sigma^2}} \right )^{-1},
\end{equation}
whereas for $\sigma^2 > 1.218$  inequality (\ref{ine2}) holds and then  $\tilde{\eta}_c^0=1$.   In Fig.\ \ref{fig:5} we have 
$\sigma^2 =1$ so that $\tilde{\eta}_c^0=2  (1 - \sqrt{3}/2)/(1-\sqrt{2}/2) \approx 0.915$.

Equation (\ref{expan}) also allows us  to determine how $\tilde{w}$, which can be seen as the order parameter of our model,\cite{Huang_1963} vanishes when $\eta$ approaches $\eta_c^0$ from below, viz.,
\begin{equation}
\tilde{w} \approx \mathcal{C} \left ( \eta_c^0 - \eta \right )
\end{equation}
where
\begin{equation}
 \mathcal{C} = \frac{\gamma}{4} \left [ \frac{\mathbb{E}^2_S(S)}{\mathbb{E}_S(S^2)} - b_2 \frac{  \mathbb{E}_S(S^3) \mathbb{E}_S(S)}{\mathbb{E}_S^2(S^2)} \right ]^{-1} ,  
\end{equation}
with $b_2$ given by Eq.~(\ref{bi}) with $\eta$ replaced by $\eta_c^0$ and we have used Eq.~(\ref{b1c}) to write $b_1$ in terms of the moments of $S$.  Hence the optimal copying probability vanishes linearly with the distance $| \eta_c^0 - \eta |$ to the critical point.

The analysis of the critical point $\eta_c^1$,  at which the regime of totally credulous individuals is replaced by a regime of skeptical individuals, is more involved and has to be done numerically. Let us begin by looking at the continuous transition exhibited in Fig.~\ref{fig:4}
for $\gamma =0.2$.  The region close to $w=1$ is amplified in  Fig.~ \ref{fig:6} revealing the   mechanism behind  that continuous transition:  it happens when the derivative  of $ \langle \Lambda^{(\infty)}(w) \rangle  $ calculated at $w=1$ vanishes. Unfortunately, we cannot study analytically $ \langle \Lambda^{(t)} (w) \rangle  $ in the vicinity of $w=1$ by, say, expanding it in powers of $1-w$. The reason is that all generations $t=0, 1, \ldots$ contribute to  the  coefficients of the powers $1-w$, $(1-w)^2$, etc. and since we do not have an explicit expression for $ \langle \Lambda^{(t)} (w) \rangle  $ for large $t$ we cannot calculate those coefficients. Nevertheless, we can easily calculate numerically the derivative of $ \langle \Lambda^{(t)} (w) \rangle  $ at $w=1$,\cite{Press_1992} provided we compute 
 $ \langle \Lambda^{(\infty)} (w) \rangle  $ with sufficiently high  precision, as done in  Fig.~ \ref{fig:6}. This is  the situation where we need to iterate the recursion equation (\ref{rec}) up to $t=10^4$ generations to produce a very precise estimate of $ \langle \Lambda^{(\infty)} (w) \rangle $.

\begin{figure}[]
\centerline{\includegraphics[width=.6\textwidth]{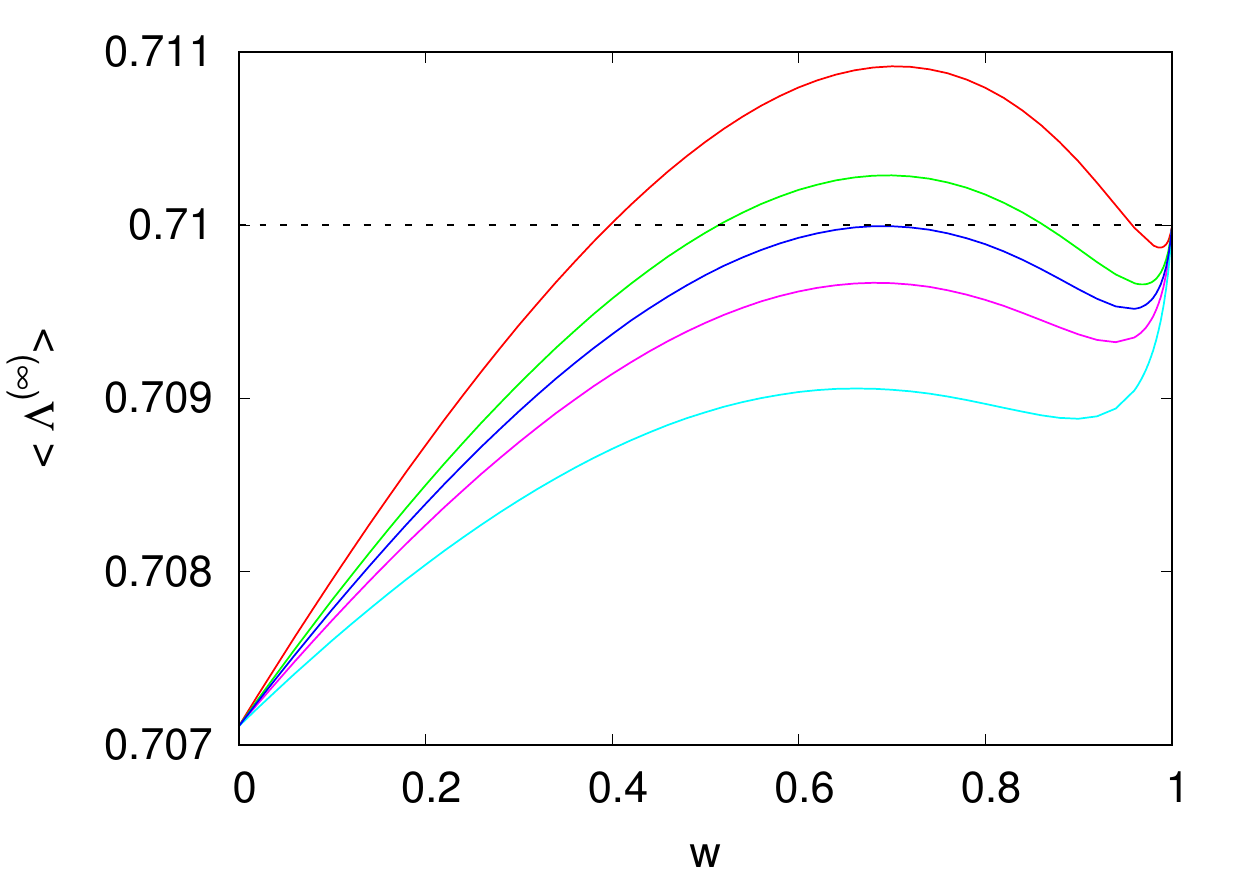}}
\vspace*{8pt}
\caption{Mean fraction of individuals that survive the environmental challenge  for a population of infinite size at equilibrium against 
the copying probability $w$ for $\gamma =0.29$, $\sigma^2 = 1$ and (top to bottom) $\eta =085, 0.86, 0.8647, 0.87 $ and $0.88$.
 The dashed horizontal  line is  $ \langle \Lambda^{(\infty)}(1) \rangle =  1-\gamma  =0.71$.  The point at which the discontinuous  transition takes place is $\eta_c^1 \approx 0.8647$.
  }
\label{fig:7}
\end{figure}

Increase of $\gamma$  leads to a distinct scenario, which is depicted in Fig.~ \ref{fig:7}  for $\gamma=0.29$.  This figure explains the discontinuous transition shown in Fig.~ \ref{fig:5}:   as $\eta$ increases the interior maximum $ \langle \Lambda^{(\infty)}( \tilde{w}) \rangle  $ decreases and eventually equals  $ \langle \Lambda^{(\infty)} (1) \rangle  $ at $\eta = \eta_c^1$.  As before,  since $\tilde{w}$ can be  evaluated numerically,  the critical point $\eta_c^1$ can be obtained rather straightforwardly. Moreover, the  size of the discontinuity vanishes when  $\tilde{w}$ tends to $1$, so the numerical evaluation of $\tilde{w}$ allows us to determine the tricritical point at which the continuous transition becomes discontinuous.\cite{Landau_1980}

\begin{figure}[]
\centerline{\includegraphics[width=.6\textwidth]{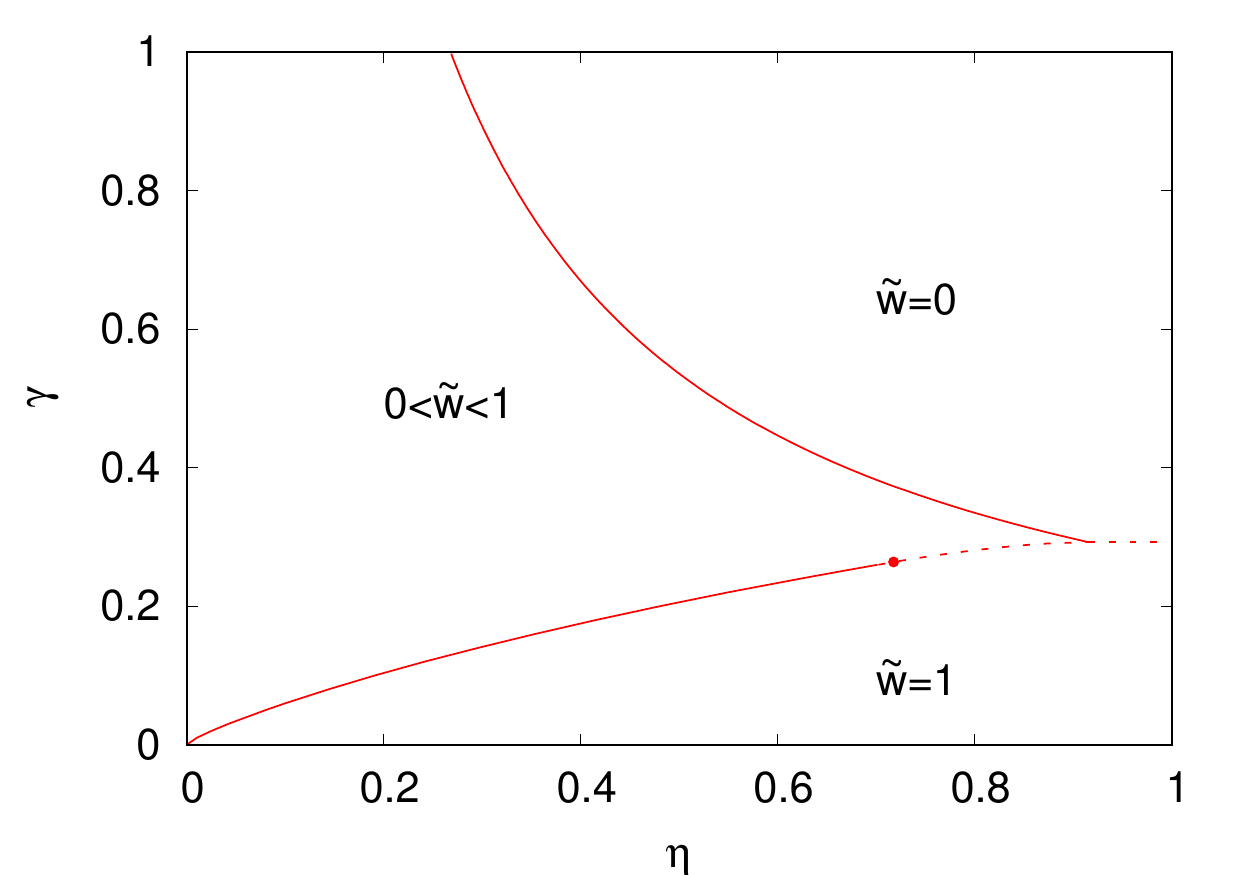}}
\vspace*{8pt}
\caption{Phase diagram  in the space $(\eta, \gamma)$ for $\sigma^2 = 1$ showing the three equilibrium regimes characterized by  distinct optimal copying strategies $\tilde{w}$.  The solid curves indicate the  continuous phase transitions whereas the 
dashed curves the discontinuous transitions. The filled circle indicates the approximate  location of the tricritical point $(0.718, 0.264)$ that  joins smoothly  the continuous and the discontinuous transition lines. The triple point is located at $(0.915, 0.293)$.
  }
\label{fig:8}
\end{figure}

Figure \ref{fig:8} exhibits the rich phase diagram for $\sigma^2 = 1$ that summarizes the results presented up to now. In addition to the continuous and discontinuous phase transitions,
we note  the existence of a triple point at which the three regimes (or phases) coexist. The triple point is determined by the conditions   $ \langle \Lambda^{(\infty)} (0) \rangle  = \langle \Lambda^{(\infty)} ( \tilde{w}) \rangle  = \langle \Lambda^{(\infty)} (1) \rangle $, which imply 
$\gamma_{triple} = 1-1/\sqrt{1+\sigma^2} $ and $\eta_{triple} = \tilde{\eta}_c^0$ with $\tilde{\eta}_c^0$ given by Eq.~(\ref{eta_max}).

To conclude our analysis, we stress that increase of the hazardousness of the environment $\sigma^2$ decreases the region of dominance of the unwelcome regime $\tilde{w} =0$ in the parameter space  $(\eta, \gamma)$. In particular, that regime disappears altogether when the minimum value of $\eta_c^0$, which is obtained by setting $\gamma =1$ in Eq.~(\ref{etac}),   is greater than  $1$ (see Fig.~\ref{fig:8}), i.e., for $\sigma^2 > 3 + 2 \sqrt{3} \approx 6.46$. Hence, for highly hazardous environments,  the exploration-only  pure strategy is never optimal, regardless of  the degree of deceitfulness  of the individuals or  of  the cost of believing  distorted information.

\section{Discussion}\label{sec:disc}

Our analytically solvable model for the effects of disinformation  on  epistemic security  offers two main lessons:
\begin{itemize}
\item   If the cost of believing corrupted information is not  low, which happens to be the case for information regarding Covid-19 as the hospitalization
and death  rates among the unvaccinated are much higher than among  the vaccinated,\cite{Dyer_2021}
then there is a maximum degree of deceitfulness the population can bear before trust is completely eroded and the  optimal  strategy for survival  becomes to trust no one. In the context of the Covid-19 pandemic, this phenomenon is observed in the slow pace of   vaccination rates among the elderly in China,\cite{Wang_2021} which  may be in part because misinformation about mRNA technology  led to mistrust of vaccines in general, but  it may also reflect a broader distrust of the government. 

\item A completely credulous population can be  surprisingly robust against  the deceitful behavior of its members,  provided the  risk of exploring the environment (i.e., of finding the truth by oneself) is not too low (see Fig.~\ref{fig:8}). In fact, copying (uncorrupted) information is an almost sure bet, since the copyist  acquires a viability that has already  passed at least one environmental challenge.  Hence our model predicts that the harsher the environment is,  the greater the trust on the information exhibited by the older generations.

\end{itemize}

In order to study our model analytically we have opted for a population-centred approach,\cite{Birch_2018} in which the quantity to be maximized is the mean fitness of the population, with no reference to the interests of their members,
which we choose to  behave identically on the average, anyway. A more plausible scenario, however, is  one where the individuals behave so as to maximize their own chances of survival and of passing their offspring to the next generation.  Of course, the latter goal is best achieved by deceiving other individuals, so they are likely to fail the environmental challenge. We have run extensive individual-based simulations of a scenario where the individuals  are assigned random values of  the   copying probability $w$,  but  the same degree of deceitfulness $\gamma$,  in the initial generation.  The game evolutionary dynamics was followed until fixation occurs, i.e., until all individuals exhibit the same value of $w$ -- the winner of the contest --   so they share  a common ancestor at some previous generation.\cite{Serva_2005}  We found that the average of the winner values of $w$ over distinct runs exhibits a  behavior that is qualitatively  similar to that  of the  population-centred quantity $\tilde{w}$ studied here. In particular, the critical point $\eta_c^0$ coincides with our theoretical estimate (\ref{etac}).  These simulation results, together with those where the initial population is heterogeneous regarding  both $w$ and $\gamma$, will be presented elsewhere.

To conclude, we note that use of evolutionary game theory to study  the  antagonism between truth-telling and  lying has a long tradition both in the biological and the  sociological contexts.\cite{Sober_1994}  More recently, however, analogous problems have been successfully addressed  by the  active particles methods, which   stand as an alternative  general strategy toward the modeling
of the collective dynamics of large systems of interacting living entities.\cite{Bellomo_2021}

\bigskip

\acknowledgments
JFF is indebted to Mauro Santos for the many discussions on social information.
The research of JFF was  supported in part 
 by Grant No.\  2020/03041-3, Fun\-da\-\c{c}\~ao de Amparo \`a Pesquisa do Estado de S\~ao Paulo 
(FAPESP) and  by Grant No.\ 305620/2021-5, Conselho Nacional de Desenvolvimento 
Cient\'{\i}\-fi\-co e Tecnol\'ogico (CNPq).
HAT was supported by the Coordena\c{c}\~ao de Aperfei\c{c}oamento de Pessoal de
N\'{\i}vel Superior - Brasil (CAPES) - Finance Code 001.


\end{document}